# A wireless triboelectric nanogenerator

Sai Sunil Kumar Mallineni[a], Yongchang Dong[a], Herbert Behlow[a], Apparao M. Rao[a,*],

Ramakrishna Podila[a,b,*],

[a] Clemson Nanomaterials Institute, Department of Physics and Astronomy, Clemson University, Clemson, SC 29634, USA.
[b] Laboratory of Nano-biophysics and COMSET, Clemson University, Clemson, SC 29634, USA.

*Corresponding author: rpodila@g.clemson.edu, arao@g.clemson.edu, Phone: 864-656-4447, Fax: 864-656-0805.

**Abstract:** We demonstrate a new paradigm for the wireless harvesting of mechanical energy via a 3D-printed triboelectric nanogenerator (TENG) which comprises a graphene polylactic acid (gPLA) nanocomposite and Teflon. The synergistic combination of eco-friendly PLA with graphene in our TENG exhibited an output voltage > 2 kV with an instantaneous peak power of 70 mW, which in turn generated a strong electric field to enable the wireless transmission of harvested energy over a distance of 3 m. Specifically, we demonstrate wireless and secure actuatation of smart-home applications such as smart tint windows, temperature sensors, liquid crystal displays, and security alarms either with a single or a specific user-defined passcode of mechanical pulses (e.g., Fibonacci sequence). Notably, such high electric output of a gPLA-based TENG enabled unprecedented wireless transmission of harvested mechanical energy into a capacitor, thus obviating the need for additional electronics or energy sources. The scalable additive manufacturing approach for gPLA-based TENGs, along with their high electrical output can revolutionize the present method of harnessing the mechanical energy available in our environment.

**1. Introduction**

Triboelectricity is emerging as a possible power source for portable electronics [1-6], sensors [7–18], and other wearable devices [19–24]. Triboelectric nanogenerators (TENGs) harness the contact induced electrostatic potential generated across the surfaces of two dissimilar materials to convert waste mechanical energy into usable electrical energy. Given that many materials such as metals, silk, wool exhibit triboelectrification, the choice of electrode materials in TENGs is virtually unlimited [25,26]. The materials pair in a TENG is often chosen so as to maximize the potential drop while allowing easy flow of charges (i.e., less electrical resistance) to harvest usable power. In the last five years, a series of proof-of-concept studies has demonstrated TENGs using pairs of different patterned nanomaterials and polymers. Notwithstanding this progress, an important question remains unaddressed in TENGs. How to realize eco-friendly and high-performance TENGs without the need for the devices to be hardwired to the TENG? It is thus imperative to identify earth-abundant, biodegradable, and recyclable materials (e.g., biopolymers) that are suitable for realizing sustainable and eco-friendly TENGs with high output electric fields for wireless transmission of harvested energy.

The crystallographic symmetry is critical in determining the tribo- and piezo-electrical properties of materials [27]. For example, the piezoelectric tensor vanishes for any material or crystal with a centre of symmetry, implying that tribo- and piezo-electrification is inadequate for effectively polarizing centrosymmetric crystals. Using crystal symmetry, Fukada et al. established that effective polarization could be achieved in biopolymers when polar groups are linked to one of their asymmetric carbon atoms [28,29]. Polylactic acid (PLA) [30], which is a plant-derived biodegradable linear aliphatic thermoplastic polyester, contains two asymmetric carbon atoms that facilitate a high degree of polarization upon tribo-electrification (see supplementary Fig. S1) [31]. Unfortunately, the high electrical resistance makes PLA unsuitable as a TENG electrode. Here, we resolve this challenge by using electrically conducting graphene-PLA (gPLA) nanocomposites to additively manufacture sustainable TENG electrodes with high output voltages (>2 kV) and high output powers (> 70 mW). Graphene is an ideal filler for improving the electrical conducting properties of PLA because it: 1) can store injected electrical charges with a decay time ~40 mins,[30] which is an order of magnitude higher than decay times in oxides, 2) leads to high electrical conductivity (volume resistivity ~0.6 ohm-cm) at low filler content ~15 wt.%, and 3) improves the mechanical robustness of PLA [32].

In this article, we describe novel additively manufactured gPLA nanocomposite-based high-performance TENGs that not only convert mechanical energy into electricity but also wirelessly (W-TENG) transmit the generated energy without the need for either additional circuitry or external electrical power. A 3D-printed gPLA nanocomposite on a polyimide (or Kapton) film was used with a complementary polytetrafluoroethylene (PTFE or Teflon) sheet to fabricate a gPLA-based TENG. When actuated by simple mechanical motions such as hand tapping, the W-TENG generated high output voltage (> 2 kV) and peak power (> 70 mW at 10 MΩ). An estimated force from hand tapping was ~ 120 N (see supplementary information) and was applied at an average frequency of ~ 3 Hz to activate W-TENG. Furthermore, the high output voltage, which resulted in a high electric field at the end of the copper ribbon (attached to the gPLA electrode, Fig. S2) was effective in enabling wireless transmission of the electric field over a distance of 3 m. Unlike conventional earlier studies in which a TENG was hardwired to power a commercial wireless transmitter [33,34], a W-TENG can wirelessly control a variety of electronic gadgets (e.g., electrochromic windows, temperature sensors, liquid crystal displays, and security alarms for smart-home applications) in real time, obviating the need for additional either amplification or commercial wireless transmitters. Unlike state-of-the-art wireless transmitters with external power systems (e.g., through batteries), W-TENGs represent a renewable self-powered alternative that can activate an electronic circuit by simple mechanical motion such as hand tapping. Lastly, we also demonstrate that the electrical energy generated from mechanical energy imparted to a W-TENG can be wirelessly transmitted and stored in a capacitor. All the above attributes make W-TENG a viable green alternative for wirelessly powering the internet of things (IoT).

2. Materials and methods

*2.1 Construction of a W-TENG*

A Prusa i3 3D printer was used for additively manufacturing TENGs using gPLA filaments purchased from Graphene Supermarket. For comparison, a similar TENG was manufactured using PLA filaments. A borosilicate heat-print-bed glass maintained at 70 °C was used as the bottom supporting substrate (Fig. 1). A thin polyimide film was first attached to the top surface of the bed glass, followed by extrusion of the PLA or gPLA filament at 220 °C and layer-by-layer printing on the polyimide film through the fused deposition model. Due to the characteristic that PLA and gPLA have poor affinity for glass, a buffer sheet of polyimide was intentionally used in the 3D printing process, otherwise the printed features would warp and peel off the bottom bed glass substrate. A copper ribbon was attached to the gPLA electrode to serve as a wireless transmitter, and a Teflon sheet (purchased from ePlastics) having a thickness of 0.25 mm was used as the top electrode (Fig. S2).

*2.2 Characterization of a W-TENG*

Micro-Raman spectroscopy was performed on the gPLA electrodes using a Renishaw micro-inVia spectrometer (514.5 nm Ar+ ion laser excitation, 50× objective and Peltier-cooled

CCD). Scanning electron microscopy (SEM, Hitachi S4800) and thermogravimetric analysis (TGA, Q500 system from TA instruments, in flowing nitrogen) were also performed, and the output TENG voltages were measured using a Yokogawa DL 9710L digital oscilloscope.

*2.3 Wireless signal processing circuit (WSPC)*

A WSPC was designed in-house, a description of which is provided later in Fig. 6, was used for the wireless detection of TENG output signals. Our WSPC consists of a preamplifier (LMC6001), an intermediate amplifier (TL062), and a pulse-shaping integrated chip or IC (NE555). A high pass filter with a 150pF series capacitor (needed to mitigate the interference from the surrounding electric fields) and a 100 MΩ resistor (characteristic roll-off frequency of ~5 Hz) was used as the high impedance input to the preamplifier. The preamp was configured with a gain of ~2.2. Although TENGs produce both negative and positive pulses upon pressing and releasing, the amplitude of the positive voltage pulse in our case was ~4-fold larger than the negative pulse. Thus, only the positive pulse was retained from the preamp output, which was passed through a Si-diode for signal rectification. The intermediate amplifier was configured as an inverting amplifier with unity gain to make the rectified signal compatible with the pulse shaping IC's trigger input. Finally, 555 timer the pulse shaping IC (see Fig. S3) was configured to operate in a one-shot monostable mode, which upon being triggered produces a 12 V square pulse of ~0.2 s duration (a signal compatible with the toggling relay trigger input). The 0.2 s duration of this one-shot output eliminates any input pulse "bounce" (from the oscillation of TENG electrode after mechanical activation) that might be present in the time window of 0.2 s. The output duration of the pulse from the 555 timer can be adjusted by modifying the values of the capacitor and resistor connected in series between pins 1 and 8. When the negative trigger pulse from the inverting amplifier is applied to pin 2 of the 555 timer, the voltage across the capacitor (4.7 μF attached to 39 KΩ; RC ~0.2 s) increases exponentially for a period of ~0.2 s. Subsequently, the output drops to a "low" as depicted in Fig. S3. Thus, the TENGs in this study were designed to transmit wireless signals with a minimum spacing of ~0.2 s.

3. Results and discussion

In this study, gPLA feedstock was heated above its glass transition temperature (Tg= 55 °C) and extruded through the 3D printer nozzle (Fig. 1a) to rapidly print multiple gPLA layers (~16 x 18 cm$^2$) on a thin polyimide (or Kapton) film (thickness ~60 μm) attached to a borosilicate heat-print-bed glass. This assembly constitutes the bottom electrode for the TENG. Narrow strips of Kapton tape were then used to attach a Cu ribbon to the printed gPLA, and a Teflon sheet to the bottom electrode to yield a W-TENG (Fig. 1c). The high electronegativity was the rationale for using Teflon, which can readily accept electrons when rubbed against other surfaces [25]. In the W-TENG depicted in Fig. 1c, the buckling of the top Teflon sheet resulted in a natural air gap (~1 mm) between the top and bottom electrodes obviating the need for additional spacers that are often used in vertical TENGs.

As shown in Fig. 2a, the Raman spectrum of the gPLA electrode showed the characteristic graphitic, or G-band (~1585 cm$^{-1}$), along with the disorder, or D-band (~1350 cm$^{-1}$), and its overtone 2D band (~2700 cm$^{-1}$).[35,36] Note the evidence of the CH$_3$ symmetric stretching modes of PLA ~2900 cm$^{-1}$ in addition to the Raman features of graphene [37]. A Thermogravimetric analysis (TGA) of the PLA and gPLA electrodes showed a clear decrease in weight at temperatures ~270 °C and ~340 °C respectively, due to the decomposition of PLA (Fig. 2b). The presence of graphene in the PLA matrix clearly increased the structural stability of the gPLA electrode. Similar enhancements in the structural composition were observed with the addition of carbon nanotubes (CNTs) into the PLA polymer matrix [32]. Unlike PLA electrodes, gPLA electrodes showed ~15-17 % weight retention above 400 °C due to the presence of graphene, which was confirmed by the Raman spectrum of gPLA electrode subjected to 800 °C during TGA (Fig. S4).

The W-TENG is initially in a neutral state with no potential difference across the electrodes. The top electrode is negatively charged when it is "pressed" against the bottom electrode by a mechanical force, such as hand tapping (see Fig. 3a-b). The top surface of gPLA is oxidized leading to a surface polarization [38,39]. Upon releasing the mechanical force, the negatively charged Teflon sheet relaxes to its initial configuration, and further polarizes the bottom gPLA electrode leading to a measurable mean potential difference > 1.5 kV (Fig. 3c-e). Such enhanced output voltages were not observed when the bottom electrode was printed using a PLA filament (Fig. S5). While the surface dipoles on PLA become oriented under the influence of negatively charged Teflon, the dipoles within its bulk remain randomly oriented due to the lack of charge flow and hindered mobility of polymer macromolecules (Fig. 4a) [38].

The voltage increase in gPLA electrode based TENG is due to the presence of graphene, which extends the PLA polarization deeper into the bulk by facilitating charge flow (see Fig. 4b). Similar enhancement in TENG performance was observed upon addition of reduced graphene oxide (rGO) in polyimide composite. Such enhancement has been attributed to additional charge trapping sites created by graphene in the dielectric matrix [40]. To further confirm this assertion, we etched the gPLA electrode surface using dicholorethane to remove the top layer of PLA on the surface of the electrode. The $CH_3$ stretching modes ~2900 $cm^{-1}$ which were present in the as-printed gPLA electrode, were absent in the Raman spectrum of dichloroethane-treated gPLA electrode (Fig. S6), thus confirming the removal of the PLA from the surface and exposure of the graphene. The W-TENGs with dichloroethane-treated gPLA electrodes, however, showed ~1.8 kV that is ~33 % lower than the voltage exhibited by as-printed gPLA electrode (2.7 kV, Fig. S7 and S11).

A detailed electrical characterization of the W-TENGs hardwired to varying loads is presented in the supplementary information (Fig. S8a). More importantly, although no significant current was drawn from the PLA electrodes, the improved electrical conductivity of gPLA electrodes facilitated a current flow with a peak power ~70 mW (Fig. S8b). The high electrical output of the W-TENG readily powered ~300 commercial green LEDs (Figs. 5a and 5c) and also rapidly charged a 10 µF capacitor to ~30 V within 2 minutes (Figs. 5b and 5d). Note that the focus of this study is to demonstrate the use of W-TENGs in self-powered wireless applications. One notable by-product here was a determination that the output characteristics depicted in Fig. S8 are superior to the characteristics of TENGs reported in the literature [6].

Clearly, the high electric field generated by the W-TENG supplants the old wireless transmission model that requires an external signal transmitter. Any mechanical action or pulse placing the top Teflon sheet in contact with the bottom gPLA electrode generated a large potential difference (> 2 kV at the device) with an associated electric field instantaneously sensed over a distance of ~3 m. The gentle hand tapping of the W-TENG was detected in real-time as a single voltage pulse by an oscilloscope equipped with a custom-built WSPC (Figs. 6a and 6b) situated ~3 m from the W-TENG. Unlike previous TENG demonstrations in which the TENG was merely used to charge batteries or capacitors to power commercial wireless signal transmitters our W-TENG is unique in that it acts both as the electrical energy generator and the signal transmitter [33,34]. When the W-TENG was hand tapped in a Fibonacci sequence (i.e., 1, 1, 2, 3, 5, and 8 taps) with a ~1 s gap between each cycle, the mechanical pulses were wirelessly detected by the WSPC as an instantaneous voltage spike with the same periodicity as the input pulses (Fig. 6c and supplementary video S1). Such a real-time response allows the self-powered W-TENGs to wirelessly transmit signals (akin to Morse coding) for detection via simple and inexpensive electronic receivers. Thus, the W-TENGs, which function as self-powered wireless controllers are useful in smart-home applications (e.g. lights, temperature sensors, burglar alarms, smart-windows, and garage doors). As shown in Fig. 7, we hand tap W-TENGs to wirelessly activate alarms/calling bells, lights, sensor displays, smart-windows, and photo frames (see supplementary videos S2-S6). Lastly, W-TENGs can be used to activate security systems with either a single, or a specific user-defined passcode via mechanical pulses (e.g., Fibonacci sequence).

Given that the most abundant energy associated with humans is mechanical energy resulting from body motion, W-TENGs can be used to harvest this otherwise wasted mechanical energy (e.g., walking) to wirelessly charge energy storage devices (e.g. capacitors). As a proof-of-concept, a 1 µF capacitor was wirelessly charged to 5.0 V within a minute (corresponding to a power of ~0.2 µW) using a W-TENG that was triggered by hand tapping (Fig. 8). Though the harvested power may seem low, this charging is 100% wireless and requires no batteries. Thus, one could envision a large arrays of W-TENGs integrated into walkways, roads and other public spaces to wirelessly charge energy storage devices that can harvest this wasted mechanical energy. Given that mechanically robust W-TENGs can be scalably 3D printed and virtually last forever, such large installations are physically feasible and economically viable. We note that the already high output of a W-TENG can be further enhanced to 3 kV via Ar plasma treatment of the top Teflon electrode [40], or by patterning or texturing the bottom gPLA electrodes via 3D printing (Fig. S9 and S10). This enhanced friction will convert mechanical energy into electricity, and wirelessly transmit energy into storage devices (e.g., capacitor).

4. Conclusion

Fused deposition modelling was used for the additively manufacturing or 3D printing of PLA-based TENGs on a polyimide film. The addition of graphene filler to PLA improved the electrical conductivity of the printed gPLA electrode, which improved the W-TENG performance with output voltages > 2 kV and output powers of ~70 mW. The high electrical output of W-TENGs readily powered ~300 commercial green LEDs and also rapidly charged a 10 µF capacitor to ~30 V within 2 minutes. The high voltage output of W-TENGs generated strong electric fields enabling wireless transmission without any external signal transmitters. In this regard, W-TENGs represent ideal self-powered transmitters for securely actuating smart-home applications (e.g. lights, temperature sensors, burglar alarms, smart-windows, and garage doors) upon receiving a specific sequence of mechanical pulses (i.e., a secure passcode). W-TENGs also permit the unprecedented wireless harvesting of mechanical energy, viz., a 1 µF capacitor wirelessly charged to 5.0 V within a minute using a W-TENG triggered by hand tapping.

**Acknowledgements**

R.P. and A.M.R are thankful to Watt Family Innovation Center (2301812) for the financial support. R. P. also thanks Clemson University for the start-up funds.

**Appendix A. Supplementary information**